# Intrinsic Redshifts in Normal Spiral Galaxies

David G. Russell

Owego Free Academy, Owego, NY 13827 USA

Russeld1@oagw.stier.org

## Abstract

The Tully-Fisher (TF) relation calibrated in both the B-band and the I-band indicates that (1) the redshift distribution of Virgo Cluster spirals has a morphological dependence that is inconsistent with a peculiar velocity interpretation. (2) Galaxies of morphology similar to ScI galaxies have a systematic excess redshift component relative to the redshift expected from a Hubble Constant of 72 km s$^{-1}$ Mpc$^{-1}$. (3) Pairs and groups of galaxies exist for which the TF relation provides excellent agreement among individual members, but for which the group redshift deviates strongly from the predictions of the Hubble Relation. It is again found that morphology plays a role as these galaxies are all of Hubble types Sbc and Sc. The overall results of this study indicate that normal Sbc and Sc galaxies have a systematic excess redshift component relative to the predictions of the standard Hubble relation assuming a Hubble Constant of 72 km s$^{-1}$ Mpc$^{-1}$. The excess redshifts identified in this analysis are consistent with the expectations of previous claims for non-cosmological (intrinsic) redshifts.

# 1. Introduction

Strong empirical evidence has accumulated that many quasars are not at large cosmological distances as is expected from the traditional redshift-distance relation (Arp 1986; Arp 1998a; Chu et al 1998; Arp 1999; Bell 2002; Lopez-Coredoira&Gutierrez 2002). The emerging picture is that quasars can be ejected from active Seyfert galaxies as high redshift objects that evolve to lower redshifts as they age (Arp 1998b, Bell 2002). Recently, Lopez-Corredoira&Gutierrez (2002) demonstrated that a pair of high z objects are present in a luminous filament apparently connecting the Seyfert galaxy NGC 7603 to the companion galaxy NGC 7603B which is previously know to have a discordant redshift. It was noted by Arp (1980) that the NGC 7603/NGC 7603B pair is one example of a class of objects in which a lower redshift galaxy appears to be physically connected by a luminous bridge to a companion with a discordant high redshift. Associations between high and low redshift companions are generally dismissed as accidental alignments. As stated by Arp (1980) "But in the end, no matter how convincing the connection is, it would be possible to take the position that it is an accident."

The most thorough investigations into the existence of non-velocity redshifts in normal galaxies have been presented by Arp (1987, 1988, 1990,1994,1998a,b) and Russell (2002). Arp demonstrated that Sbc/Sc type galaxies have excess redshifts relative to Sb galaxies in the Virgo cluster (1988) and in the general field (1990). Russell (2002) utilized linear diameters as a further test of this discrepancy and confirmed this behavior.

The Hubble Key Project (Freedman et al 2001) has provided an increased number of calibrators for the Tully-Fisher relation. Russell (2002) demonstrated that scatter in the Tully-Fisher relation can be reduced when morphological type dependence is properly accounted for in calibrating the Tully-Fisher relationship. Using type dependent Tully-Fisher relations in both the B and the I-bands it is shown that ScI galaxies and a number of small groups of galaxies possess sizeable systematic excess redshifts which suggest the existence of an intrinsic redshift component.

## 2. Type dependent Tully-Fisher relations

### 2.1 Calibration of type dependent Tully-Fisher relations

Russell (2002 – R02) demonstrated that the Tully-Fisher relation is subject to morphological and luminosity class type dependence such that galaxies of morphology similar to ScI galaxies and Seyfert galaxies are more luminous at a given rotational velocity than Sab/Sb galaxies and Sc galaxies of luminosity classes II-III to IV (figure 1).  In this analysis, two morphological groups are identified for calibration of the Tully-Fisher relation.   The ScI group includes SbcI,SbcI-II,ScI,ScI-II, and Seyfert galaxies(filled circles in figure 1).  The Sb/ScIII group includes all Sab/Sb/Sbc/Sc galaxies that are not in the ScI group.   For this analysis, the type dependent equations have been calibrated using corrected B-band and I-band magnitudes from the LEDA database following the calibration procedures described in R02.   The Tully-Fisher(TF) calibrator data is provided in Table I and Table II gives the B-band and I-band zero points.  Figure 1 is a plot of the TF relation in the B-band and the  B-band  and I-band TF relations are:

B-band:

$\mu$   =   19.85(+/- 0.16) + 5.24(+/- 0.10)(log Vmax - 2.2) + Btc   Sb/ScIII group

$\mu$   =   20.44(+/-0.09) + 4.91(+/-0.20) (log Vmax - 2.2) + Btc    ScI group

I-band:

$\mu$   =   21.29(+/-0.24) + 7.12(+/-0.19)(log Vmax - 2.2) + Itc     Sb/ScIII group

$\mu$   =   21.55(+/-0.07) + 7.22(+/-0.10)(log Vmax - 2.2) + Itc     ScI group

The major concern with the procedure adopted here for calibration of the TF relation is the smaller number of calibrators available for slope and zero point determination when the calibrators are split into type dependent samples.  However, a few points lessen this concern.  First, the slopes of the Sb/ScIII and ScI type dependent relations are very close in both the B-band and the I-band.  If the complete calibrator samples were used for determining the slope, then the slopes would be 5.75 for the B-band relation and 7.39 for the I-band relation.   Second, the overall scatter is smaller with the use of the type dependent equations.  In the B-band the 1$\sigma$ scatter relative to the calibrator distance moduli is 0.09 mag for the ScI group and 0.16 mag for the Sb/ScIII group. In the I-band the $1\sigma$ scatter is 0.07 mag and 0.24 mag respectively. With a single calibration the Hubble Key Project (Sakai et al 2000) found TF scatter of 0.45 mag (B-band) and 0.37 mag (I-band) for a sample of 21 cepheid calibrators. Finally, it is noted that Tully&Pierce (2000) found that the slope and zero point derived from their cluster template relations were virtually identical to the slope and zero point derived from the calibrator sample alone. This suggests that useful TF relations can be derived from the calibrator samples.

If the type effect is not accounted for, ScI group galaxies will have systematically underestimated TF distances while Sb/ScIII group galaxies will have systematically overestimated TF distances. This effect is illustrated with a sample drawn from the Ursa Major cluster sample of Verheijen (2001) which are listed in Table III. Rotational velocities utilized for calculating TF distances for this sample are Vflat from Verheijen (2001). With the use of Type dependent equations, the mean distance modulus of the ScI group galaxies agrees within 0.04 mag of the mean distance modulus to the cluster. However, when a single calibration is utilized the mean distance modulus of the ScI group is less than the mean distance modulus of the Sb/ScIII group sample by 0.45 mag (B-band) and 0.27 mag (I-band).

## 2.2 The Tully-Fisher Sample

For this analysis we have selected (1) Galaxies within the Virgo cluster with TF data in the LEDA database. (2) Galaxies of ScI group morphology (SbcI,SbcI-II, ScI, ScI-II) and rotational velocities derived from optical rotation curves in Mathewson&Ford (1996) or rotational velocities derived from hydrogen linewidths in Giovanelli et al (1997). (3) Small groups and pairs of galaxies with rotational velocities in Mathewson/Ford. ScI group galaxies selected were restricted to those with inclinations between 30 and 80 degrees and with absorption corrections in LEDA (ag + ai) less than 1.00 mag. Galaxies in the ScI group were also restricted to those with mean surface brightness in the $D_{25}$ isophote (bri25 in LEDA) from 22.35 to 23.99 – which covers the surface brightness range of the calibrators. To reduce the effect of inclination errors upon corrected magnitudes, galaxies were eliminated for which the LEDA inclination has a difference from the Mathewson/Ford or Giovanelli et al inclination that exceeds 7 degrees. After these restrictions, the sample size for the ScI group was 83 galaxies.

# 3.  Evidence for non-cosmological redshifts

## 3.1 The Virgo Cluster

The Virgo cluster is the nearest major cluster for which evidence for intrinsic redshifts in normal galaxies has previously been indicated (Arp 1988).   Extensive data for TF analysis is available in the LEDA database and the recent surface brightness fluctuation study of E/SO galaxies of Tonry et al (2001) provides additional distances.   In Table IV, the mean distances and redshift velocities of subsamples of Virgo cluster galaxies are provided.  While all subsamples have mean distances within 1.4 Mpc of the mean distance for the full sample of 77 galaxies, two outstanding anomalies emerge in Table IV.   First, there are dramatic differences in mean redshift that are dependent upon morphological type.  Second, galaxies with linear diameters less than 20 kpc within the E/SO group and galaxies with rotational velocities less than log v =2.200 within the SO-a to Sbc group have significantly larger mean redshifts than the giant galaxies within the same morphological groups.

The most dramatic result in Table IV is the extreme excess redshifts of the ScI/Seyfert group. Since three of these galaxies (NGC 4321, NGC 4535, NGC 4536) have Cepheid distances it is unlikely that this phenomenon results from inaccurate distances (see also Arp 2002).  The result cannot be attributed to the morphological density relation (Dressler 1980) because the redshift excess is systematically positive and the galaxies in question are on both the front and backside of the mean cluster distance.

The unusual redshift-morphology result is further illustrated in Table V for which the distances, redshifts and peculiar motions (calculated with Vvir in LEDA assuming $H_0$=72 km s$^{-1}$ Mpc$^{-1}$) are given for all galaxies within the core distance of 13-20 Mpc and with linear diameters exceeding 20 kpc.  Note that the giant Sab/Sb galaxies have apparent large negative peculiar motions in contrast to the large positive apparent peculiar motions of the ScI/Seyfert group.  *Adopting a strict velocity interpretation of galaxy redshifts requires that as a group the giant Sb galaxies are approaching the Milky Way with a mean velocity of -898 km s$^{-1}$ while the giant ScI galaxies are receding from the Milky Way with a mean velocity of +824 km s$^{-1}$.*

The second discrepant result in Table IV is the excess redshift that galaxies smaller than 20 kpc have relative to the giant galaxies in the E/SO and Sb groups.  As with morphologically selected subsamples, the diameter selected subsamples are close to the mean cluster distance.  However, relative to the larger galaxies in each group, the small galaxies have a systematic excess redshift of 417 km s$^{-1}$ for the elliptical group and 277 km s$^{-1}$ for the Sb group.   This excess extends to the dwarf elliptical and spheroidal galaxies in Virgo (Binggeli, Popescu, & Tammann 1993). Similar excesses for smaller companions have also been demonstrated for the Local Group and M-81 group (Arp 1988,1994).

The morphological dichotomy of redshifts in the large spirals of Virgo is incompatible with a peculiar velocity interpretation. If peculiar motions were responsible for deviations from the velocities predicted by the Hubble relation, then as many negative as positive peculiar motions should be expected in both groups. Instead the trend is clearly systematic with Sb galaxies systematically deficient in redshift and ScI galaxies systematically excessive in redshift. But this exactly what is seen in the Local Group and M-81 group for which the largest Sb type spirals have the smallest redshifts(Arp 1994).

An alternative explanation is that the excess redshifts of the ScI group are non-velocity or intrinsic. A model that may account for both the excess redshift of the ScI group and the redshift deficit of the Sb group will be noted in section 4.

## 3.2. ScI galaxies

Table VI lists the 83 ScI group galaxies utilized in this analysis. In figures 2 and 3 the Hubble relation for these galaxies is plotted for the B-band and I-band TF relations respectively. The solid line in figures 2 and 3 represents a Hubble Constant of 72 km s$^{-1}$ Mpc$^{-1}$ (Freedman et al 2001). It is clearly seen that the ScI galaxies have a systematic excess redshift relative to a Hubble Constant of 72 km s$^{-1}$ Mpc$^{-1}$. Since the result is the same for both the B-band and the I-band TF relations it is unlikely that magnitude errors are the source of the problem. It is also important to note that the use of type dependent TF equations is not the reason for the difference between redshift and TF distances either. The ScI group TF zero point is 0.59 mag larger than the Sb/ScIII group zero point in the B-band and 0.26 mag larger in the I-band. *If a single TF calibration for all types was utilized, the TF distance moduli of the ScI group galaxies would be reduced and thus would make the discrepancy even more extreme.*

The bulk of the sample has rotational velocities derived from optical rotation curves by Mathewson&Ford (1996 - MF). Since the MF rotation curves were demonstrated to have an internal consistency of 10 km s$^{-1}$ or better it is unlikely that rotational velocity errors are the source of the deviation from the predicted Hubble relation. It should also be noted that if rotational velocity errors are the explanation, then the MF optical rotation curves systematically underestimate the true rotational velocity.  However, Verheijen (2001) has noted that the velocity at which the rotation curve becomes flat may be a better estimate of the true rotational velocity than the maximum rotational velocity of a rotation curve.  So it would not be unexpected that many of the ScI galaxies in our sample have overestimated rotational velocities and TF distances too large.  Correcting for these errors would make the mean deviation from the predicted Hubble relation even more extreme.

   If it is assumed that there are no intrinsic redshifts, then the ScI group sample utilized here indicates a Hubble constant of 86 km s$^{-1}$ Mpc$^{-1}$(B-band) or 85 km s$^{-1}$ Mpc$^{-1}$(I-band).  The systematic excess redshift of ScI galaxies was previously demonstrated by Arp (1990) with the B-band TF relation for a large sample of ScI type galaxies.   The results of this study reinforce and improve upon the earlier result of Arp in several ways:  (1) Both B and I-band TF relations have been utilized and the TF relation is calibrated from a larger number of Cepheid calibrators.  The 1$\sigma$ scatter of the relation is less than 0.10 mag for both the B-band and I-band among the calibrators and no ScI group calibrator has a TF distance modulus that differs from the Cepheid distance modulus by more than 0.15 mag. (B-band) or 0.24 mag. (I-band).   (2) Optical rotation curve rotational velocities were utilized in this study while Arp (1990) utilized hydrogen linewidths and thus both methods of measuring rotational velocity produce the same result.   (3) The sample utilized here is drawn independently of Arp's sample and extends to larger distances and higher redshifts.

   In a number of cases the individual discrepancy between the TF distance and the H$_0$=72 redshift distance is significantly larger than can be accounted for by TF errors and therefore strongly supports the existence of intrinsic redshifts.   For example, the galaxy ESO 147-5 has a TF distance modulus of 34.65 (B-band) or 34.56 (I-band) while the redshift distance modulus is 35.86 – a discrepancy of 1.21 to 1.30 mag.  ESO 147-5 thus has an excess redshift of +4551 km s$^{-1}$.  An excess redshift this large cannot be a peculiar motion or result from a large scale flow and therefore indicates the existence of a non-cosmological (intrinsic) redshift component.   In the next section it is shown that this interpretation is supported by a number of pairs and small groups of galaxies.

## 3.3 Pairs and Groups with intrinsic redshifts

   Table VII lists TF distance moduli for eight pairs/groups of galaxies with rotational velocities in Mathewson & Ford (1996).     Several important points must be made regarding these groups.  (1) Four of the groups include ScI galaxies with extreme excess redshifts that were discussed in the previous section.  (2) The scatter in TF distance moduli among the galaxies in these pairs/groups is consistent with the small TF scatter of the calibrators. (3) The galaxies in these groups are all (except the edge-on galaxy ESO 243-8) types Sbc or Sc as classified in LEDA. (4) The groups exhibit extreme excess redshifts comparable with the most extreme examples among the ScI group galaxies as is demonstrated in Table VIII.

The close agreement of the TF distance moduli for the galaxies in these eight pairs/groups makes it unlikely that the TF distances suffer from significant errors and therefore these groups verify the existence of intrinsic redshifts indicated by the ScI group galaxies. The Hubble relation for these pairs and groups is plotted in figure 4. It can be seen that the mean redshifts of these groups deviate from the predicted Hubble relation by 1021 km s$^{-1}$ to 4746 km s$^{-1}$. For example, the ESO 549-30 group includes 6 galaxies with a mean TF distance modulus of 34.22. The redshift distance modulus indicated by the group mean velocity is 35.27 and requires that either all 6 galaxies have TF distance modulus errors of ~ 1.00 mag or that the galaxies in the group have a mean intrinsic redshift of +3140 km s$^{-1}$.

## 4. Discussion and Conclusion

The results of this analysis support the previous conclusions of Arp (1988, 1990, 1998a) that many late type spiral galaxies contain an excess redshift component that may reasonably be interpreted as non-cosmological or intrinsic. It was also found that the giant Sab/Sb galaxies in Virgo have extreme negative peculiar motions or intrinsic redshifts. Arp (1988 – and references therein) noted that the redshift deficit of Sb galaxies relative to other morphological types in clusters may be a common phenomenon. A distant cluster of galaxies in Serpens with TF data in the LEDA database was identified which also demonstrates this behavior. The individual galaxies are listed in Table IX and the group is plotted in figure 5. It can be seen in figure 5 that the galaxies in this group form a filament with the Sa galaxy PGC 55828 located near the center of the filament (indicated by the filled triangle in figure 5). The mean distance of the group is 154.2 Mpc and the mean redshift is 10835 km s$^{-1}$. Thus this group indicates a Hubble Constant of 70.3 km s$^{-1}$ Mpc$^{-1}$ which is consistent with the results of Freedman et al (2001). However, the redshift-morphology behavior of the Virgo cluster is repeated in this distant group as shown in Table X. The mean distance of the Sa/Sb and Sbc/Sc subsamples are in excellent agreement, but the mean redshift of the Sbc/Sc galaxies is 1507 km s$^{-1}$ greater than the Sa/Sb galaxies. This difference would increase to 1748 km s$^{-1}$ if the single Sb group galaxy with logvmrad <2.200 is eliminated from the Sb group sample. In addition, individual galaxies exist within the Serpens filament that have extreme deviations from the mean cluster redshift. The SBbc galaxy PGC 55724 has a redshift of 15065 which is 4230 km s[-1] greater than the cluster mean. In constrast, the central Sa galaxy PGC 55828 has a redshift of 6989 km s[-1] which is 3846 km s[-1] less than the cluster mean. In the traditional interpretation of galaxy redshifts PGC 55828 is a foreground galaxy. But in that view it would have to be accepted that PGC 55828 has an incorrect TF distance modulus that by chance matches the TF distance moduli of 16 other galaxies forming a real group. At the same time it would have to be accepted that PGC 55724 and PGC 56169 are a pair of background galaxies which have incorrect TF distance moduli that also by chance match the TF distance moduli of 16 other galaxies in a real group. It is also important to note that both PGC 55828 and PGC 55724 exhibit normal double-horned hydrogen profiles in Freudling, Haynes, & Giovanelli (1992) – which reinforces that large TF errors are unlikely to be the reason for the difference between the redshift and TF distance moduli.

Narlikar&Arp (1993) provided a model which may explain the intrinsic redshift behavior identified in this analysis. In the model of Narlikar&Arp (1993) galaxy redshifts are a function of age such that younger galaxies have larger redshifts than older galaxies at the same distance. The results of this analysis suggest that late type Sbc/Sc galaxies tend to have larger redshifts than early type Sa/Sb galaxies. In the Narlikar&Arp model this would indicate that late type spirals are generally younger than early type spirals. For PGC 55828, the low redshift leads to the interpretation that it is the oldest galaxy in the filament according to the Narlikar&Arp model.

Regardless of the theoretical explanation for intrinsic redshifts, the empirical results of this study strongly indicate that intrinsic redshifts exist in normal galaxies. Tully-Fisher distances in both the B-band and the I-band have been used to establish that ScI group galaxies have a systematic excess redshift relative to redshifts predicted with a Hubble Constant of 72 km s[-1] Mpc[-1]. These excess redshifts must be even more extreme if proponents of a lower Hubble Constant are correct (eg. Ekholm et al 1999; Parodi et al 2000; Sandage 2002). In the most extreme cases the difference between redshift and Tully-Fisher distances implies redshift excesses larger than 3000 km s[-1] for groups and up to +4550 km s[-1] for individual galaxies.

## Acknowledgements:

This research has made use of the Lyon-Meudon extragalactic Database (LEDA) compiled by the LEDA team at the CRAL-Observatoire de Lyon (France).

Table I:  Tully-Fisher Calibrator Data

| | 1 | 2 | 3 | 4 | 5 | 6 | 7 | 8 |
|---|---|---|---|---|---|---|---|---|
| Galaxy | Type | btc | Itc | incl | logVrot | m-Mcalib | Ref. |
| Sb/ScIII group | | | | | | | |
| N224 | SbI-II | 3.20 | 1.47 | 78 | 2.411 | 24.48 | 1 |
| N2841 | SbI | 9.51 | 7.91 | 67 | 2.488 | 30.74 | 2 |
| N3031 | SabI-II | 7.12 | 5.41 | 59 | 2.375 | 27.80 | 1 |
| N3351 | SBbII | 10.06 | 8.32 | 48 | 2.232 | 30.00 | 1 |
| N3368 | SBabII | 9.75 | 7.99 | 49 | 2.336 | 30.11 | 1 |
| N4527 | SBbcII | 10.62 | 8.91 | 74 | 2.260 | 30.75 | 3 |
| N4548 | SBbI-II | 10.66 | 8.89 | 38 | 2.279 | 31.05 | 1 |
| N4639 | SBbcII-III | 11.81 | 10.27 | 54 | 2.215 | 31.71 | 1 |
| N4725 | SBabI.3 | 9.75 | 8.04 | 64 | 2.325 | 30.46 | 1 |
| N598 | ScII-III | 5.73 | | 55 | 2.011 | 24.62 | 1 |
| N2090 | ScII-III | 10.94 | 9.35 | 69 | 2.146 | 30.35 | 1 |
| N2403 | SBcIII | 8.24 | 7.32 | 59 | 2.104 | 27.54 | 1 |
| N2541 | SBcIII-IV | 11.61 | 10.53 | 63 | 1.989 | 30.25 | 1 |
| N3319 | SBcII | 11.33 | 10.39 | 57 | 2.043 | 30.62 | 1 |
| N4414 | ScII-III | 10.65 | 8.87 | 57 | 2.344 | 31.24 | 1 |
| | | | | | | | |
| ScI group | | | | | | | |
| N1365 | SBbI(sy) | 9.90 | 8.18 | 47 | 2.397 | 31.27 | 1 |
| N1425 | SbII.2 | 10.83 | 9.55 | 65 | 2.270 | 31.70 | 1 |
| N2903 | SBbcI-II | 8.86 | 7.58 | 64 | 2.282 | 29.75 | 4 |
| N3198 | SBcII | 10.22 | 9.22 | 68 | 2.184 | 30.70 | 1 |
| N3627 | SBbII(sy) | 8.98 | 7.54 | 54 | 2.326 | 30.01 | 1 |
| N4258 | SBbcII-III(sy) | 8.38 | 7.04 | 72 | 2.327 | 29.51 | 1 |
| N4321 | ScI | 9.79 | 8.39 | 30 | 2.338 | 30.91 | 1 |
| N4535 | SBcI-II | 10.35 | 8.89 | 43 | 2.270 | 30.99 | 1 |
| N4536 | SBbcI-II | 10.43 | 9.15 | 68 | 2.230 | 30.87 | 1 |
| N4603 | SBcI.3 | 11.35 | 9.86 | 49 | 2.358 | 32.61 | 5 |
| N5457 | ScdI | 8.24 | 6.99 | 21 | 2.315 | 29.13 | 1 |
| N7331 | SbcI-II | 9.25 | 7.70 | 67 | 2.424 | 30.84 | 1 |

Table II:  Calibration of the B-band and I-band Tully-Fisher Relations

| | 1 | 2 | 3 | 4 | 5 | 6 | 7 |
|---|---|---|---|---|---|---|---|
| Galaxy | zero pt B | m-M $TF_B$ | m-Merr B | zero pt I | m-M $TF_I$ | m-Merr I | |
| Sb/ScIII group | | | | | | | |
| N224 | 20.17 | 24.16 | -0.32 | 21.51 | 24.26 | -0.22 | |
| N2841 | 19.72 | 30.87 | 0.13 | 20.78 | 31.25 | 0.51 | |
| N3031 | 19.76 | 27.71 | -0.09 | 21.14 | 27.95 | 0.15 | |
| N3351 | 19.77 | 30.08 | 0.08 | 21.45 | 29.84 | -0.16 | |
| N3368 | 19.65 | 30.31 | 0.20 | 21.15 | 30.25 | 0.14 | |
| N4527 | 19.82 | 30.78 | 0.03 | 21.41 | 30.63 | -0.12 | |
| N4548 | 19.98 | 30.92 | -0.13 | 21.60 | 30.74 | -0.31 | |
| N4639 | 19.82 | 31.74 | 0.03 | 21.33 | 31.67 | -0.04 | |
| N4725 | 20.06 | 30.26 | -0.20 | 21.53 | 30.22 | -0.24 | |
| N598 | 19.88 | 24.59 | -0.03 | | | | |
| N2090 | 19.69 | 30.51 | 0.16 | 21.38 | 30.26 | -0.09 | |
| N2403 | 19.80 | 27.59 | 0.05 | 20.90 | 27.93 | 0.39 | |
| N2541 | 19.75 | 30.35 | 0.10 | 21.22 | 30.32 | 0.07 | |
| N3319 | 20.11 | 30.36 | -0.26 | 21.35 | 30.56 | -0.06 | |
| N4414 | 19.84 | 31.25 | 0.01 | 21.34 | 31.25 | -0.05 | |
| | | | $\sigma$=0.16 | | | $\sigma$=0.24 | |
| ScI group | | | | | | | |
| N1365 | 20.40 | 31.31 | 0.04 | 21.67 | 31.15 | -0.12 | |
| N1425 | 20.53 | 31.61 | -0.09 | 21.64 | 31.61 | -0.09 | |
| N2903 | 20.49 | 29.70 | -0.05 | 21.58 | 29.72 | -0.03 | |
| N3198 | 20.56 | 30.58 | -0.12 | 21.60 | 30.65 | -0.05 | |
| N3627 | 20.41 | 30.04 | 0.03 | 21.56 | 30.00 | -0.01 | |
| N4258 | 20.51 | 29.44 | -0.07 | 21.55 | 29.51 | 0.00 | |
| N4321 | 20.44 | 30.91 | 0.00 | 21.52 | 30.94 | 0.03 | |
| N4535 | 20.30 | 31.13 | 0.14 | 21.59 | 30.95 | -0.04 | |
| N4536 | 20.29 | 31.02 | 0.15 | 21.50 | 30.92 | 0.05 | |
| N4603 | 20.48 | 32.57 | -0.04 | 21.61 | 32.55 | -0.06 | |
| N5457 | 20.33 | 29.24 | 0.11 | 21.31 | 29.37 | 0.24 | |
| N7331 | 20.40 | 30.78 | -0.06 | 21.52 | 30.87 | 0.03 | |
| | | | $\sigma$=0.09 | | | $\sigma$=0.07 | |

Table III: The effect of Type dependence in the Ursa Major cluster

| Galaxy | Type | Log Vflat | m-M B Type | m-M B Single | m-M I Type | m-M I Single |
|---|---|---|---|---|---|---|
| NGC 3726 | SBcI-II | 2.210 | 31.02 | 30.65 | 30.96 | 30.82 |
| NGC 3893 | SBcI.2 | 2.274 | 31.62 | 31.31 | 31.53 | 31.41 |
| NGC 3953 | SBbcI | 2.348 | 31.39 | 31.13 | 31.37 | 31.25 |
| NGC 3992 | SBbcI | 2.384 | 31.68 | 31.46 | 31.58 | 31.47 |
| NGC 4051 | SBbc II $_{SY}$ | 2.201 | 31.20 | 30.83 | 30.86 | 30.72 |
| NGC 4100 | SbcI-II | 2.215 | 31.46 | 31.10 | 31.27 | 31.13 |
| NGC 3769 | SBbII-III | 2.086 | 31.01 | 31.16 | 31.19 | 31.28 |
| NGC 3949 | SbcIII-IV | 2.215 | 31.16 | 31.38 | 31.42 | 31.54 |
| NGC 3972 | SBbcIII-IV | 2.127 | 31.53 | 31.70 | 31.71 | 31.81 |
| NGC 4102 | SBbII | 2.250 | 31.68 | 31.92 | 31.43 | 31.56 |
| NGC 4217 | SbIII | 2.250 | 31.38 | 31.62 | 31.04 | 31.17 |
| NGC 4389 | SBbcIV | 2.041 | 31.21 | 31.34 | 30.85 | 30.92 |
| NGC 3729 | SBaIII-IV | 2.179 | 31.60 | 31.80 | 31.44 | 31.55 |
| NGC 4138 | SO-a | 2.167 | 31.78 | 31.97 | 31.12 | 31.23 |
| NGC 4088 | SBcII-III | 2.238 | 30.43 | 30.66 | 30.72 | 30.85 |
| NGC 4085 | SBcIII-IV | 2.127 | 31.59 | 31.76 | 31.54 | 31.64 |
| NGC 3917 | ScIII-IV | 2.130 | 30.79 | 31.05 | 31.09 | 31.19 |
| UGC 6983 | SBcIII | 2.029 | 31.96 | 32.09 | 31.96 | 32.04 |
| | | | | | | |
| Mean All | | | 31.36 | 31.38 | 31.27 | 31.31 |
| Mean Sb/Sc | | | 31.34 | 31.53 | 31.29 | 31.40 |
| Mean ScI | | | 31.40 | 31.08 | 31.26 | 31.13 |

Table IV: Virgo cluster Hubble Constant

| Type | Sample size | Mpc | Vvir | $H_0$ |
|---|---|---|---|---|
| Elliptical | 27 | 17.1 | 1062 | 62.1 |
| >20 kpc | 9 | 16.8 | 784 | 46.7 |
| <20 kpc | 18 | 17.2 | 1201 | 69.8 |
| | | | | |
| SO-a to Sbc | 21 | 18.4 | 1010 | 54.9 |
| Log V > 2.200 | 10 | 17.9 | 865 | 48.3 |
| Log V< 2.200 | 11 | 18.8 | 1142 | 60.7 |
| | | | | |
| ScII-III to ScIV | 18 | 18.2 | 1310 | 72.0 |
| | | | | |
| Sbc and Sc I/I-II, seyferts | 11 | 19.4 | 1854 | 95.6 |
| | | | | |
| All | 77 | 18.0 | 1226 | 68.1 |

Table V: Peculiar motions of Virgo Cluster galaxies larger than 20 kpc

| Galaxy | Type | Mpc | Vvir | PV$_{72}$ |
|---|---|---|---|---|
| NGC 4254 | ScI.3 | 17.0 | 2505 | +1281 |
| NGC 4303 | SBbcI | 16.1 | 1624 | +465 |
| NGC 4321 | SBbcI | 15.2 | 1682 | +588 |
| NGC 4501 | SbI-II Seyfert | 19.9 | 2380 | +947 |
| NGC 4535 | SBcI-II | 15.8 | 1846 | +891 |
| NGC 4536 | SBbcI-II | 14.9 | 1607 | +773 |
| NGC 4527 | SbcII | 14.1 | 1776 | +761 |
| NGC 4192 | SBbII | 15.1 | -46 | -1133 |
| NGC 4216 | SBbII | 15.3 | 281 | -821 |
| NGC 4548 | SBbI-II | 16.2 | 587 | -579 |
| NGC 4569 | SBabI-II | 12.8 | -137 | -1059 |
| NGC 4365 | E/SO | 19.9 | 1299 | -134 |
| NGC 4374 | E/SO | 17.9 | 1004 | -285 |
| NGC 4406 | E/SO | 16.7 | -198 | -1400 |
| NGC 4472 | E/SO | 15.9 | 940 | -204 |
| NGC 4486 | E/SO | 15.6 | 1363 | +240 |
| NGC 4526 | E/SO | 16.4 | 518 | -663 |
| NGC 4552 | E/SO | 14.9 | 380 | -693 |
| NGC 4621 | E/SO | 17.8 | 520 | -762 |

Table VI: ScI sample

| Galaxy | incl | logVrot | Mpc B | Mpc I | Vcmb |
|--------|------|---------|-------|-------|------|
| 412-10 | 44 | 2.362 | 73.1 | 78.3 | 5790 |
| 476-15 | 58 | 2.146 | 18.5 | 18.5 | 1368 |
| 545-21 | 58 | 2.299 | 52.0 | 57.8 | 4722 |
| 299-4 | 59 | 2.267 | 55.5 | 57.8 | 5191 |
| 547-14 | 33 | 2.265 | 17.3 | 17.1 | 1577 |
| 202-26 | 54 | 2.134 | 51.3 | 44.1 | 5084 |
| IC 387 | 46 | 2.386 | 73.5 | 82.8 | 4480 |
| 553-3 | 50 | 2.185 | 56.8 | 51.8 | 4507 |
| N3029 | 48 | 2.207 | 71.8 | 73.1 | 6911 |
| 217-12 | 42 | 2.134 | 41.3 | 39.8 | 3693 |
| 323-27 | 60 | 2.316 | 52.8 | 54.5 | 4140 |
| 323-25 | 59 | 2.316 | 50.6 | 56.5 | 4518 |
| 514-10 | 39 | 2.201 | 32.1 | 30.8 | 3034 |
| 186-21 | 65 | 2.301 | 66.4 | 75.9 | 5539 |
| 74-6 | 61 | 2.344 | 30.8 | 34.3 | 3049 |
| 286-79 | 58 | 2.468 | 46.6 | 64.9 | 4751 |
| 342-43 | 63 | 2.238 | 51.7 | 49.7 | 4839 |
| 342-50 | 56 | 2.152 | 35.0 | 29.5 | 2426 |
| 404-12 | 35 | 2.260 | 48.1 | 43.1 | 2393 |
| 404-27 | 67 | 2.149 | 37.8 | 38.0 | 2278 |
| 404-31 | 65 | 2.076 | 50.8 | 51.1 | 4156 |
| 147-5 | 34 | 2.274 | 85.1 | 81.7 | 10678 |
| 534-32 | 69 | 2.292 | 89.5 | 95.1 | 8894 |
| 349-32 | 58 | 2.474 | 79.8 | 82.8 | 6534 |
| 411-10 | 41 | 2.152 | 79.4 | 67.3 | 5574 |
| 79-8 | 52 | 2.389 | 113.2 | 128.2 | 10500 |
| 541-1 | 59 | 2.324 | 71.5 | 68.9 | 6067 |
| 244-21 | 46 | 2.360 | 105.7 | 97.3 | 6983 |
| 244-43 | 60 | 2.212 | 67.6 | 68.6 | 6025 |
| 354-17 | 73 | 2.243 | 50.8 | 66.7 | 5440 |
| 52-20 | 63 | 2.196 | 63.1 | 70.8 | 8068 |
| 478-6 | 54 | 2.346 | 76.2 | 66.1 | 5103 |
| 547-31 | 49 | 2.230 | 17.8 | 14.9 | 1421 |
| 566-9 | 53 | 2.236 | 49.4 | 53.7 | 4481 |
| N3145 | 62 | 2.412 | 44.5 | 51.1 | 3998 |
| N4030 | 40 | 2.373 | 33.7 | 26.2 | 1814 |
| N5172 | 58 | 2.410 | 52.5 | 59.2 | 4306 |
| IC 900 | 48 | 2.356 | 80.2 | 90.0 | 7353 |
| 445-58 | 59 | 2.283 | 56.5 | 60.2 | 5309 |
| 186-27 | 71 | 2.301 | 35.0 | 35.2 | 2579 |
| 107-36 | 57 | 2.344 | 31.3 | 34.4 | 3003 |
| 237-2 | 66 | 2.362 | 54.5 | 58.9 | 4983 |
| N753 | 46 | 2.317 | 51.1 | 47.2 | 4644 |
| N4574 | 51 | 2.129 | 38.2 | 35.6 | 3250 |
| 383-2 | 64 | 2.262 | 64.1 | 70.8 | 6453 |
| N3726 | 53 | 2.193 | 15.4 | 14.7 | 1072 |
| N3953 | 62 | 2.331 | 18.2 | 17.8 | 1240 |
| N4100 | 77 | 2.264 | 21.9 | 21.1 | 1273 |
| 33-32 | 56 | 2.253 | 50.6 | 55.2 | 4785 |
| 564-35 | 49 | 2.057 | 10.2 | 8.6 | 1188 |
| N3672 | 69 | 2.312 | 29.5 | 28.2 | 2221 |
| 479-40 | 66 | 2.431 | 108.2 | 131.8 | 10338 |
| 186-47 | 69 | 2.210 | 60.6 | 61.4 | 4482 |
| 569-22 | 50 | 2.330 | 50.6 | 54.2 | 4091 |
| 352-33 | 42 | 2.230 | 73.8 | 68.2 | 5441 |
| 151-40 | 47 | 2.326 | 91.6 | 88.7 | 7277 |
| 114-21 | 52 | 2.241 | 94.2 | 68.2 | 6284 |
| 30-14 | 31 | 2.228 | 66.7 | 61.9 | 8207 |
| 545-13 | 34 | 2.350 | 86.7 | 91.6 | 9907 |
| 84-9 | 59 | 2.117 | 55.2 | 66.7 | 5054 |
| N2980 | 62 | 2.387 | 68.6 | 78.3 | 6050 |
| 566-26 | 47 | 2.288 | 42.3 | 47.6 | 4099 |
| N4246 | 62 | 2.267 | 55.2 | 46.8 | 4064 |
| 268-37 | 53 | 2.267 | 68.9 | 74.1 | 5175 |
| 445-27 | 60 | 2.389 | 88.3 | 121.9 | 11735 |
| 235-20 | 51 | 2.188 | 42.5 | 41.9 | 4515 |
| 108-13 | 64 | 2.127 | 39.9 | 34.5 | 2823 |
| N268 | 48 | 2.582 | 113.8 | 171.4 | 5165 |
| 304-16 | 70 | 2.312 | 69.2 | 74.5 | 5189 |
| 159-2 | 51 | 2.233 | 61.9 | 60.8 | 4318 |
| 438-15 | 65 | 2.215 | 35.7 | 42.7 | 3689 |
| N4541 | 68 | 2.396 | 75.9 | 91.6 | 7223 |
| 576-51 | 51 | 2.223 | 50.1 | 48.5 | 5501 |
| 446-58 | 60 | 2.360 | 62.2 | 61.7 | 4581 |
| 102-22 | 61 | 2.461 | 46.8 | 58.6 | 4336 |
| 234-22 | 59 | 2.314 | 56.0 | 59.7 | 5593 |
| 236-37 | 55 | 2.248 | 53.7 | 70.8 | 5384 |
| 467-27 | 64 | 2.288 | 61.1 | 53.7 | 4993 |
| 603-22 | 56 | 2.380 | 47.0 | 46.8 | 2785 |
| U4308 | 49 | 2.182 | 58.6 | 38.0 | 3787 |
| N4835 | 76 | 2.252 | 26.7 | 22.1 | 2439 |
| N5161 | 73 | 2.227 | 20.7 | 23.0 | 2660 |
| N7610 | 53 | 2.172 | 48.3 | 53.0 | 3196 |

Table VII: Pairs and Groups with intrinsic redshifts

| Galaxy | Type | RA | Dec | logVrot | Incl. | m-M(B) | m-M(I) |
|---|---|---|---|---|---|---|---|
| 52-20 | SBbcI-II | 2 05 44 | -71 06 56 | 2.196 | 63 | 34.00 | 34.25 |
| 53-2 | Sc | 2 13 12 | -70 54 49 | 2.170 | 90 | 34.01 | 34.61 |
| 30-14 | ScI-II | 2 17 57 | -76 04 50 | 2.228 | 31 | 34.12 | 33.96 |
| | | | | | | | |
| 545-13 | ScI | 2 24 40 | -19 08 28 | 2.35 | 34 | 34.69 | 34.81 |
| 545-18 | Sbc | 2 27 05 | -19 05 40 | 2.276 | 76 | 34.81 | 35.15 |
| | | | | | | | |
| 549-30 | Sbc | 3 53 54 | -17 35 36 | 2.292 | 70 | 34.08 | 34.79 |
| 549-31 | Sbc | 3 54 25 | -19 11 26 | 2.258 | 68 | 34.16 | 34.41 |
| 549-37 | SBbc | 3 56 19 | -21 49 15 | 2.201 | 51 | 34.20 | 34.74 |
| 549-40 | Sc | 3 57 10 | -18 46 42 | 2.442 | 68 | 34.61 | 34.79 |
| 550-18 | Sbc | 4 17 12 | -17 51 23 | 2.373 | 65 | 34.16 | 34.82 |
| 550-26 | Sbc | 4 21 37 | -17 55 46 | 2.230 | 84 | 34.13 | 34.57 |
| | | | | | | | |
| 31-15 | SBbc | 3 30 47 | -73 46 59 | 2.318 | 71 | 34.69 | 34.94 |
| 31-18 | ScII | 3 37 43 | -72 23 29 | 2.279 | 63 | 34.64 | 35.19 |
| | | | | | | | |
| 71-5 | SBbcII | 18 16 25 | -71 34 49 | 2.371 | 46 | 32.85 | 33.25 |
| 71-14 | SbcII-III | 18 31 07 | -71 41 33 | 2.328 | 53 | 32.69 | 33.21 |
| | | | | | | | |
| 234-13 | SbcII-III | 20 22 32 | -50 45 11 | 2.225 | 67 | 33.77 | 34.10 |
| 234-22 | SbcI-II | 20 24 22 | -50 25 58 | 2.314 | 59 | 33.74 | 33.88 |
| 234-24 | SbcII | 20 25 27 | -51 31 54 | 2.265 | 90 | 33.86 | 34.40 |
| 234-32 | SBbc | 20 28 06 | -51 41 27 | 2.233 | 62 | 33.90 | 34.18 |
| 186-47 | SbcI-II | 20 27 46 | -52 23 04 | 2.210 | 69 | 33.91 | 33.94 |
| | | | | | | | |
| 243-1 | Sbc | 00 46 07 | -42 31 55 | 2.090 | 46 | 34.40 | 34.49 |
| 295-12 | ScI | 00 50 21 | -41 14 34 | 2.294 | 55 | 34.63 | 34.62 |
| 243-3 | Sc | 00 50 34 | -43 03 53 | 2.029 | 50 | 34.39 | 34.45 |
| 243-8 | Sb | 00 53 46 | -45 11 08 | 2.215 | 90 | 34.05 | 34.21 |
| 243-14 | ScII | 00 56 59 | -45 24 41 | 2.220 | 67 | 34.29 | 34.59 |
| 243-20 | Sc | 01 03 14 | -43 00 36 | 2.037 | 77 | 33.92 | 34.34 |
| 243-25 | Sbc | 01 05 17 | -42 54 50 | 2.097 | 56 | 33.99 | 34.22 |
| 243-30 | ScII | 01 07 10 | -42 23 23 | 2.155 | 57 | 34.54 | 34.52 |
| | | | | | | | |
| 147-2 | Sc | 22 35 40 | -61 32 51 | 2.246 | 68 | 34.76 | 35.02 |
| 147-5 | SBcI-II | 22 41 37 | -57 36 18 | 2.274 | 34 | 34.65 | 34.56 |
| 147-10 | Sbc | 22 48 57 | -57 53 37 | 2.281 | 90 | 34.75 | 35.54 |

Table VIII   Excess redshifts of small groups

| Group | Vcmb | Mpc B | $PV_{72}$ B | Mpc I | $PV_{72}$ I | m-M$_{72-B}$ | m-M$_{72-I}$ |
|---|---|---|---|---|---|---|---|
| 30-14 | 8075 | 64.3 | +3445 | 71.5 | +2927 | 1.21 | .98 |
| 545-13 | 9847 | 89.1 | +3432 | 99.1 | +2712 | .93 | .70 |
| 549-40 | 8166 | 69.8 | +3140 | 86.7 | +1924 | 1.05 | .58 |
| 31-15 | 8278 | 85.9 | +2093 | 103.3 | + 840 | .63 | .23 |
| 71-5 | 3709 | 35.8 | +1131 | 44.3 | +519 | .80 | .33 |
| 234-22 | 5240 | 58.6 | +1021 | 66.1 | +481 | .47 | .21 |
| 295-12 | 7360 | 71.8 | +2191 | 78.0 | +1745 | .77 | .59 |
| 147-10 | 11075 | 87.9 | +4746 | 101.9 | +3738 | 1.21 | .89 |

Table IX: Serpens Cluster

| Galaxy | Type | RA | Dec | logVrot | Incl. | m-M(B) | Vcmb |
|--------|------|-----|-----|---------|-------|--------|------|
| 55198 | Sab | 15.480 | 25.748 | 2.342 | 75 | 35.80 | 10365 |
| 55618 | SBb | 15.618 | 25.565 | 2.459 | 55 | 36.16 | 10562 |
| 55724 | SBbc | 15.662 | 25.742 | 2.487 | 63 | 35.91 | 15065 |
| 55810 | Sc | 15.695 | 23.204 | 2.388 | 46 | 36.43 | 10477 |
| 55828 | Sa | 15.705 | 23.808 | 2.362 | 50 | 36.12 | 6989 |
| 55979 | Sc | 15.762 | 22.879 | 2.371 | 32 | 36.31 | 12378 |
| 56038 | Sc | 15.789 | 21.531 | 2.349 | 53 | 35.95 | 12651 |
| 56175 | Sbc | 15.844 | 20.382 | 2.305 | 55 | 36.23 | 11151 |
| 56227 | Sa | 15.858 | 24.435 | 2.239 | 67 | 35.87 | 9698 |
| 56169 | Sc | 15.841 | 18.139 | 2.507 | 36 | 36.06 | 14095 |
| 55721 | SBab | 15.659 | 23.199 | 2.081 | 56 | 35.95 | 11620 |
| 55708 | Sab | 15.655 | 24.455 | 2.438 | 71 | 35.85 | 10429 |
| 55380 | Sa | 15.550 | 25.569 | 2.221 | 54 | 35.78 | 10177 |
| 55872 | Sc | 15.725 | 17.313 | 2.284 | 50 | 35.59 | 9127 |
| 56532 | Sc | 15.974 | 18.039 | 2.390 | 63 | 35.74 | 12746 |
| 55530 | Sc | 15.594 | 21.503 | 2.200 | 56 | 35.65 | 12697 |
| 86655 | Sc | 15.566 | 21.790 | 2.156 | 68 | 36.46 | 7175 |
| 56186 | Sc | 15.847 | 22.239 | 2.234 | 56 | 35.72 | 9533 |
| 56163 | SBa | 15.835 | 25.055 | 2.301 | 38 | 35.69 | 9609 |
| 55213 | SBbc | 15.484 | 25.458 | 2.371 | 51 | 35.51 | 10163 |

Table X: Serpens Cluster Hubble Constant

| Type | Sa/Sab/Sb | Sbc/Sc |
|------|-----------|--------|
| Sample size | 8 | 12 |
| Mean distance modulus | 35.90 | 35.96 |
| Mean distance (Mpc) | 151.4 | 155.6 |
| Mean V3k | 9931 | 11438 |
| $H_0$ | 65.6 | 73.5 |

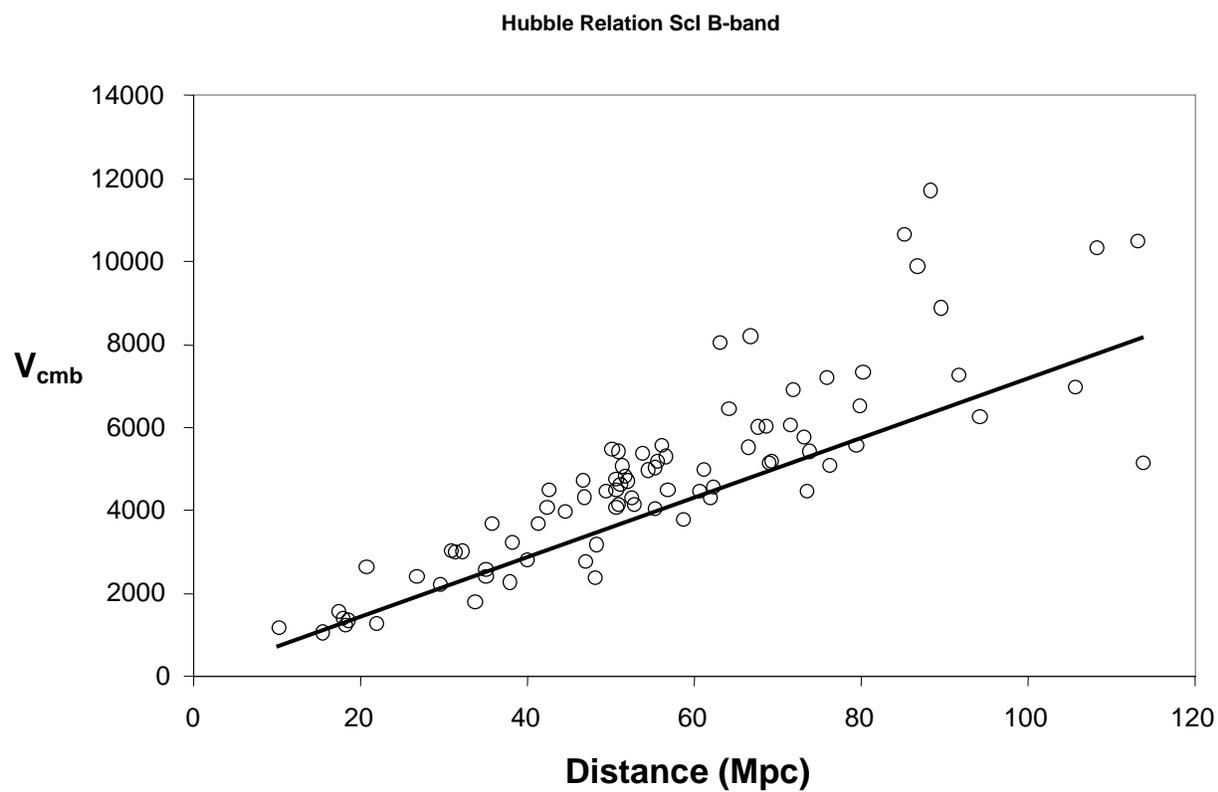

Figure 2: Hubble Relation for ScI galaxies with B-band TF relation.  Solid line is a Hubble constant of 72 km s$^{-1}$ Mpc$^{-1}$.

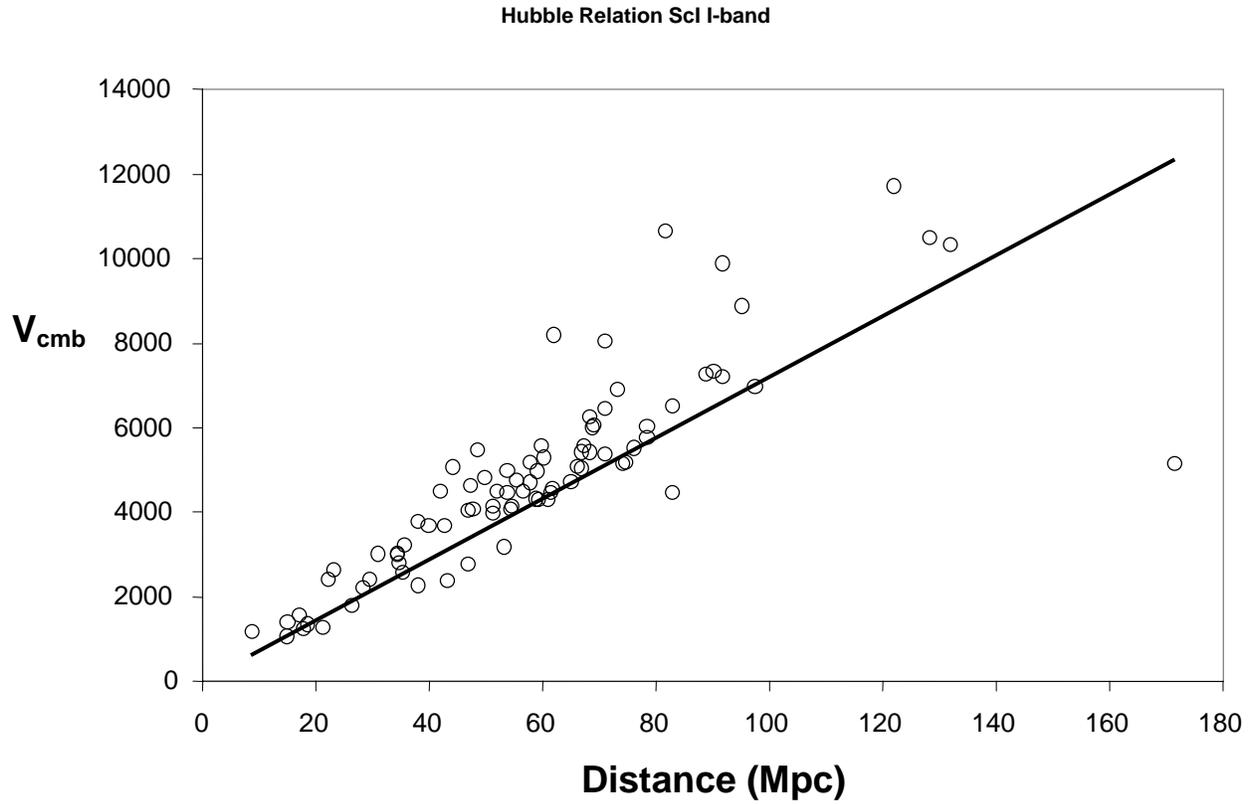

Figure 3: Hubble Relation for ScI galaxies with I-band TF relation.  Solid line is a Hubble constant of 72 km s$^{-1}$ Mpc$^{-1}$.

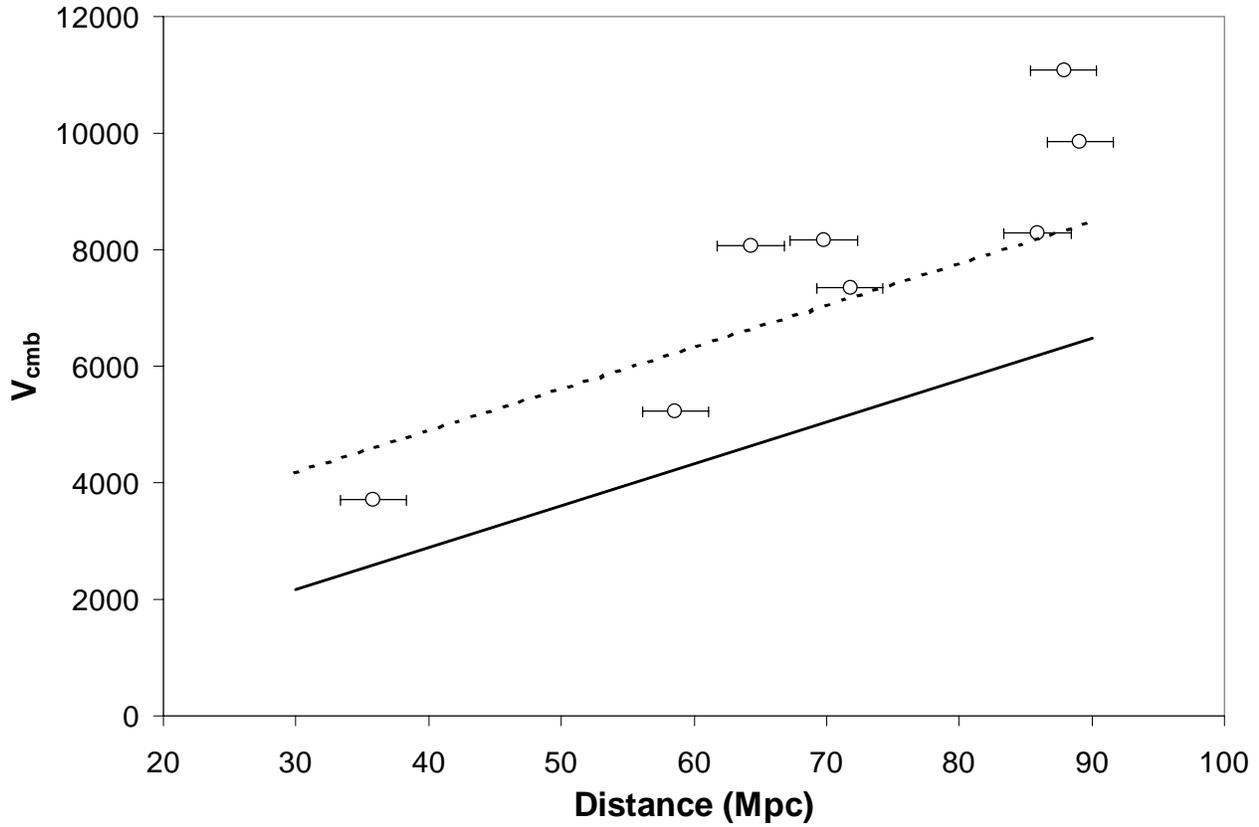

Figure 4 – Hubble Relation for pairs and groups of late type spirals in Table V. Solid line is a Hubble Constant of 72 km s$^{-1}$ Mpc$^{-1}$. Dashed line is a peculiar velocity of +2000 km s$^{-1}$. Error bars are +/- 2.5 Mpc which is typical distance error individual galaxies in the groups have from the mean group distance.

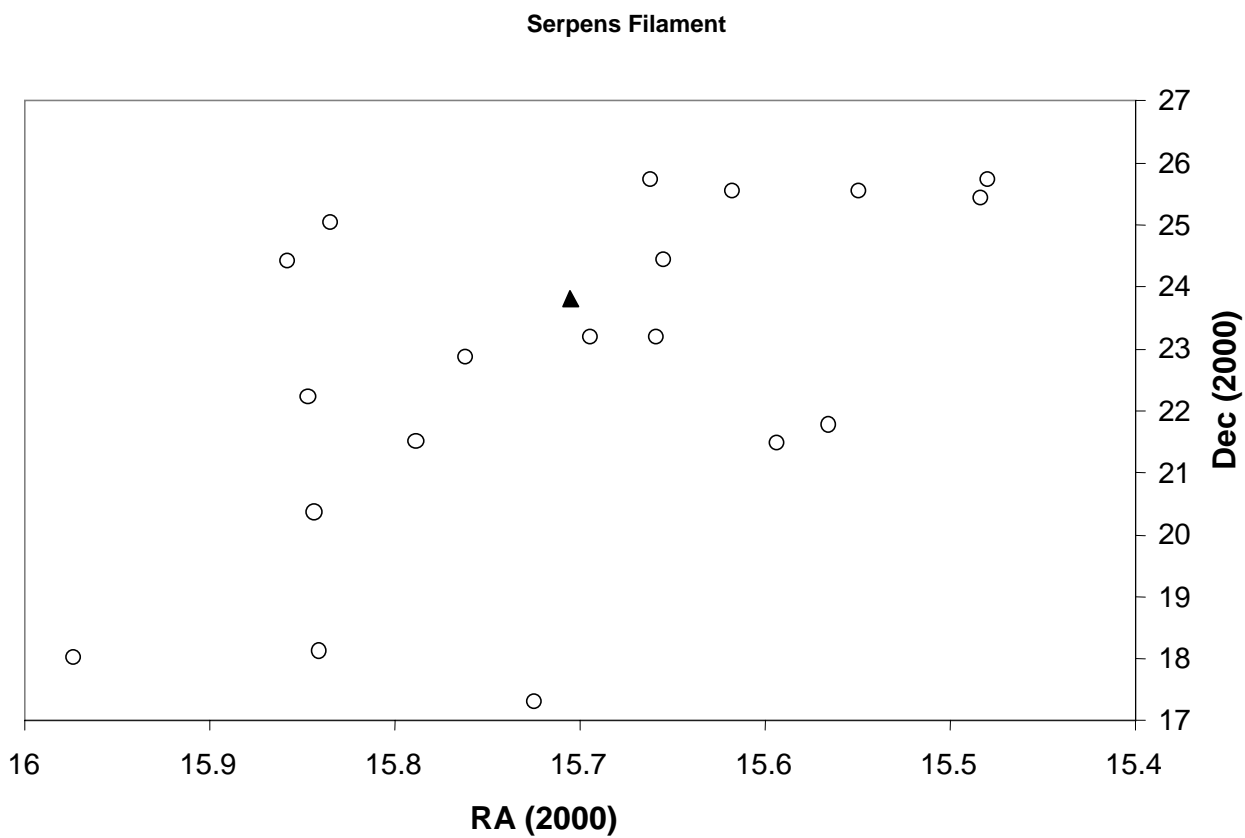

Figure 5 – Galaxies with Tully-Fisher distances which place them in the Serpens filament. The mean distance for the cluster is 154.2 Mpc. Filled triangle is the Sa galaxy PGC 55828 which has a redshift of 6989 km s[-1].

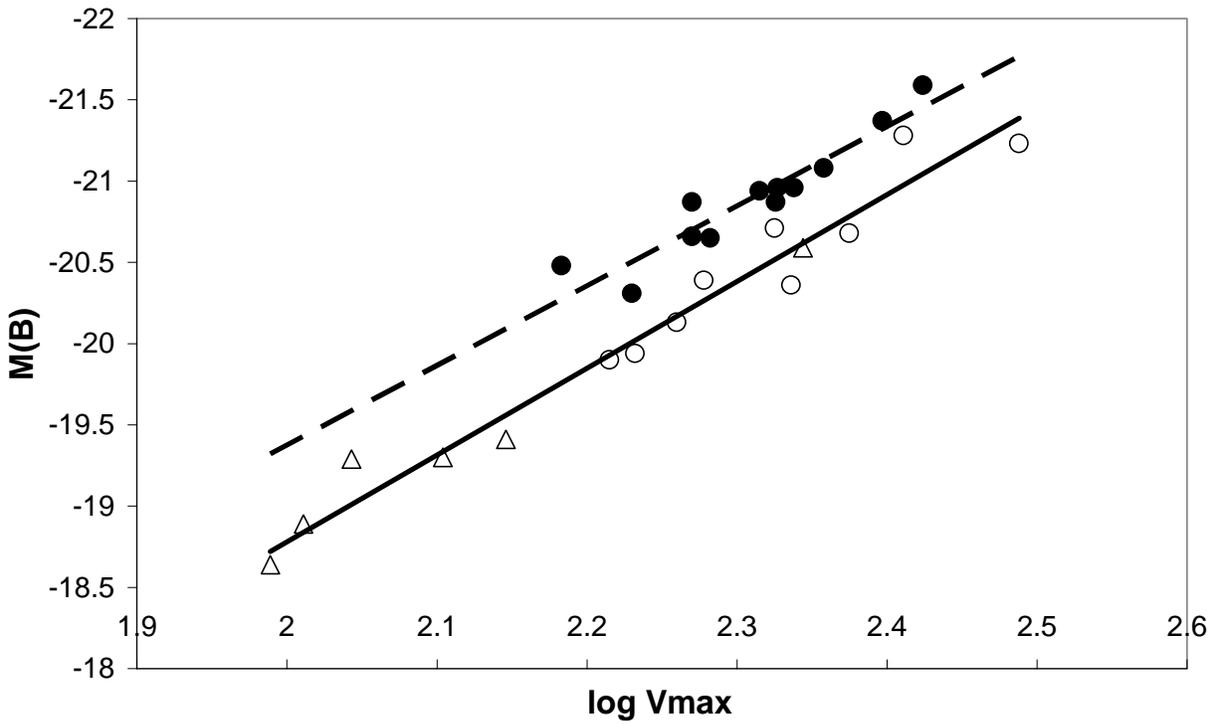

Figure 1 - B-band TF relation for TF calibrator sample.  Filled circles are ScI group galaxies. Open triangles are ScIII group galaxies.  Open circles are Sb group galaxies  Solid line is least squares fit to the Sb group.  Dashed line is least squares fit to ScI group.  Note that the ScI group galaxies are systematically more luminous at a given rotational velocity than Sb and ScIII group galaxies.